# Rhombic Fermi surfaces in a ferromagnetic MnGa thin film with perpendicular magnetic anisotropy

M. Kobayashi[1,2,*], N. H. D. Khang[3], T. Takeda[1], K. Araki[1], R. Okano[1], M. Suzuki[4], K. Kuroda[5], K. Yaji[6], K. Sugawara[7,8,9,10], S. Souma[8,9], K. Nakayama[7,10], K. Yamauchi[11], M. Kitamura[12], K. Horiba[12], A. Fujimori[4,13], T. Sato[7,8,9], S. Shin[5], M. Tanaka[1,2], and P. N. Hai[2,3,**]

[1]*Department of Electrical Engineering and Information Systems, The University of Tokyo, 7-3-1 Hongo, Bunkyo-ku, Tokyo 113-8656, Japan*
[2]*Center for Spintronic Research Network, The University of Tokyo, 7-3-1 Hongo, Bunkyo-ku, Tokyo 113-8656, Japan*
[3]*Department of Electrical and Electronic Engineering, Tokyo Institute of Technology, 2-12-1 Ookayama, Meguro, Tokyo 152-0033, Japan*
[4]*Department of Physics, The University of Tokyo, 7-3-1 Hongo, Bunkyo-ku, Tokyo 113-0033, Japan*
[5]*Institute for Solid State Physics, The University of Tokyo, 5-1-5 Kashiwanoha, Kashiwa, Chiba 277-8581, Japan*
[6]*Research Center for Advanced Measurement and Characterization, National Institute for Materials Science, Ibaraki, Japan.*
[7]*Department of Physics, Tohoku University, Sendai 980-8577, Japan*
[8]*WPI-Advanced Institute for Materials Research (WPI-AIMR), Tohoku University, Sendai 980-8577, Japan*
[9]*Center for Spintronics Research Network, Tohoku University, Sendai 980-8577, Japan*
[10]*PRESTO, JST, 4-1-8 Honcho, Kawaguchi, Saitama, 332-0012, Japan*
[11]*Institute of Scientific and Industrial Research, Osaka University, Ibaraki, Osaka 567-0047, Japan*
[12]*Photon Factory, Institute of Materials Structure Science, High Energy Accelerator Research Organization (KEK), 1-1 Oho, Tsukuba 305-0801, Japan*
[13]*Department of Applied Physics, Waseda University, Okubo, Shinjuku, Tokyo 169-8555, Japan*
\* masaki.kobayashi@ee.t.u-tokyo.ac.jp
\*\* pham.n.ab@m.titech.ac.jp

**Abstract**

Mn$_{1-x}$Ga$_x$ (MnGa) with the $L1_0$ structure is a ferromagnetic material with strong perpendicular magneto-crystalline anisotropy. Although MnGa thin films have been successfully grown epitaxially and studied for various spintronics devices, fundamental understandings of its electronic structure are still lacking. To address this issue, we have investigated $L1_0$-MnGa thin films using angle-resolved photoemission spectroscopy (ARPES). We have observed a large Fermi surface with a rhombic shape in the $k_x$-$k_y$ plane overlapping neighboring Fermi surfaces. The $k_z$ dependence of the band structure suggests that the band dispersion observed by ARPES comes from the three-dimensional band structure of MnGa folded by a $\sqrt{2} \times \sqrt{2}$ reconstruction. The band dispersion across the corner of the rhombic Fermi surface forms an electron pocket with a weak $k_z$ dependence. The effective mass and the mobility of the bands crossing the Fermi level near the corner are estimated from the ARPES images. Based on the experimental findings, the relationship between the observed band structure and the spin-dependent properties in MnGa-based heterostructures is discussed.

**Introduction**

Ferromagnetic thin films with perpendicular magnetic anisotropy (PMA) are key materials for high-density magnetic recording and various spintronics device applications. To obtain strong PMA, multilayers consisting of Co or Fe and heavy metals, such as Pt or Pd, have been usually used. The strong PMA in these multilayers originates from the strong spin-orbit coupling in the heavy metals. $Mn_{1-x}Ga_x$ thin films with the $L1_0$ (or CuAu type) crystal structure (referred to as MnGa), in contrast, contain only light metals, yet they show strong PMA with an anisotropy magnetic field as high as 40 – 50 kOe. Epitaxial MnGa thin films have been grown successfully by molecular beam epitaxy (MBE) since the early 1990s [1,2]. MnGa shows a small magnetization ($M_s$ < 400 emu/cm$^3$) [3-13] and a small damping factor ($\alpha$ < 0.008) [14], which favor low-power magnetization switching. Recently, device structures using MnGa layers have been studied, such as magnetic tunnel junctions [15], semiconductor spin-valves [16], and spin-orbit-torque devices [17]. Furthermore, remarkable spin-related phenomena, such as giant spin Hall effect [18] and giant interfacial Dzyaloshinskii-Moriya interaction (DMI) [19] have been observed in heterojunctions of the ferromagnet MnGa and the topological insulator $Bi_{1-x}Sb_x$ (BiSb). To understand the role of MnGa in these heterostructures and the physical properties of MnGa itself, it is important to characterize the electronic structure of MnGa. Previous band-structure calculations for MnGa [20,21] suggest that the density of states near the Fermi level ($E_F$) predominantly consist of the Mn 3$d$ states, and bands near $E_F$ have large exchange splitting. However, experimental studies on the electronic band structure of MnGa have not been reported so far. In this study, we have performed angle-resolved photoemission spectroscopy (ARPES) studies using vacuum ultraviolet light on an $L1_0$-MnGa thin film with PMA to elucidate its electronic states.

**Sample preparation and APRES measurements**

10 nm-thick $Mn_{0.6}Ga_{0.4}$ (MnGa) layers with the $L1_0$ crystal structure (Fig. 1(a)) were deposited on semi-insulating GaAs(001) substrates as MnGa(001)/GaAs(001) by molecular beam epitaxy (MBE) [19]. The in-plane ($a_0$) and out-of-plane lattice constants ($c$) of the primitive cell for the $L1_0$ structure are 0.272 and 0.365 nm, respectively. High crystallinity and surface morphology were confirmed by reflection high-energy electron diffraction (RHEED) during the MBE growth. The MnGa thin films were covered by amorphous Se passivation layers after the growth in order to protect the surfaces from oxidation, as shown in Fig. 1(b). Prior to the APRES measurements, the samples were heated in a preparation chamber to remove the amorphous Se layers and to obtain clean surfaces. The clean surfaces were confirmed by low-energy electron diffraction (LEED), as shown in Fig. 1(c).

Vacuum-ultraviolet (VUV) ARPES measurements using synchrotron radiation with circular polarization were performed at BL-28A of Photon Factory (PF), High Energy Accelerator Research Organization (KEK). The total energy resolution was 20 – 60 meV for ARPES measurements using photon energy ($h\nu$) of 40 – 120 eV. The ARPES measurements were conducted with a DA30 electron analyzer at 20 K under the base pressure below $1.0 \times 10^{-8}$ Pa. The binding energies were calibrated by measuring the $E_F$ of a gold foil that electrically contacted the samples. The inner potential was assumed to be ~8.0 eV. High-resolution ARPES measurements were carried out using a custom-built Scienta-Omicron DA30-L and monochromatized He-I light source ($h\nu$ = 21.2 eV) at 20 K under base pressure below $1.0 \times 10^{-8}$ Pa. The total energy resolution including thermal broadening was ~10 meV.

Band-structure calculations within density-functional theory have been performed for ferromagnetic MnGa with the $L1_0$ crystal structure. The calculations were carried out by a projector augmented wave method implemented in the Vienna ab initio Simulation Package (VASP) code [22] with the generalized gradient approximation (GGA) [23]. The cutoff energy for the plane-wave expansion of the wave functions was set to 400 eV, and a $\Gamma$-centered $12 \times 12 \times 8$ $k$-point mesh and $80 \times 80$ $k$-point mesh were used for the Brillouin-zone integration and Fermi-lines plotting, respectively.

**Results and discussion**

Figure 2(a) shows a Fermi surface mapping (FSM) of the MnGa thin film in the $k_x$-$k_y$ plane taken at $h\nu = 84$ eV, where the surface-normal momentum $k_z$ which is proportional to the square root of $h\nu$ is approximately on the $\Gamma$-X-M plane (See Figs. 1(d) and 3(a)). The FSM shows a large Fermi surface (FS) with a rhombic shape centered at the $\Gamma$ point in the $k_x$-$k_y$ plane, which reflects the symmetry of the conventional ($\sqrt{2} \times \sqrt{2}$-reconstructed) unit cell rather than the symmetry of the primitive cell. The symmetry of the FS reflects the Brillouin zone of the conventional unit cell with the lattice parameter of $\sqrt{2}a_0$ (= $a \sim 0.39$ nm) (Fig. 1(a)). The LEED pattern (Fig. 1(c)) indicates the $(\sqrt{2} \times \sqrt{2})R45°$ reconstruction of the surface of the MnGa film, consistent with the surface Brillouin zone which coincides with the surface Brillouin zone of the conventional unit cell. Hereafter, the symmetry points in the Brillouin zone of the reconstructed surface are denoted by letters with upper bars, i.e., $\overline{X}$ and $\overline{M}$. The area of the rhombic FS is approximately half of that of the reconstructed surface Brillouin zone. The FSs seem to cross each other around the $\bar{\text{X}}$ points. Figure 2(b) shows an FSM taken at $h\nu$ = 21.2 eV, where the value of $k_z$ is ~3.5 ($\pi/a$), the middle point between the $\Gamma$ and Z points in the bulk Brillouin zone of MnGa. The peak positions of the momentum distribution curve (MDC) plotted with open circles in Fig. 2(b) suggest that a small Fermi surface exists around the $\bar{\text{X}}$ point (indicated by a dashed circle), which is electron-like as discussed below. There are small differences between the FSs taken with different $h\nu$'s [for instance, the FS for $h\nu$ = 21.2 eV looks more curved than that for $h\nu$ = 82 eV (Fig. 2(a))]. This result may imply that the band structure of the MnGa film only weakly depends on $k_z$ or the $k_z$ broadening smears out the $k_z$ dependence in the ultraviolet ARPES measurements [24,25].

To examine the $k_z$ dependence of the band structure, we have measured the out-of-plane FSMs in the $k_z$-$k_x$ plane by varying $h\nu$. Figure 3(a) shows the FSM in the $k_z$-$k_x$ plane obtained by varying $h\nu$ from 50 eV to 120 eV. Figures 3(b) and 3(c) are constant-energy surface mappings (CESMs) at binding energies ($E_B$'s) of 0.3 eV and 1.1 eV, respectively. The obtained $k_z$-$k_x$ FSs are only weakly dispersive along the $k_z$ direction from $k_z$ = 5.0 to 6.0 ($\pi/c$), while the FSs are strongly dispersive above $k_z$ = 6.0 ($\pi/c$). As shown in Figs. 3(b) and 3(c), the CESMs are more strongly dependent on $k_z$, indicating that the band structure of MnGa is generally dispersive along $k_z$. The results suggest that the FSs arise basically from the three-dimensional (3D) band structure of MnGa.

Figure 4(a) shows an ARPES image taken at $h\nu$ = 82 eV along the $\Gamma$-$\bar{\text{X}}$-M line. The plot indicates that an electron-like band crosses $E_F$ around the $\bar{\text{X}}$ points. While the band dispersion seems symmetric with respect to the $\Gamma$ point, the band dispersion around the $\bar{\text{X}}$ point is not symmetric with respect to the $\bar{\text{X}}$ point. To see the band dispersion in more detail, Figs. 4(b) and 4(c) show the second derivative plot of the ARPES spectrum taken at $h\nu = 82$ eV ($k_z \sim 6.0$ ($\pi/a$)) and 21.2 eV ($k_z \sim 3.5$ ($\pi/a$)), respectively. Here, the peak positions estimated from Lorentzian fitting for MDCs are also plotted. As shown in Fig. 4(a), we have observed a multiband structure, e.g., a faster band dispersing from ~0.8 eV below $E_F$ at the Γ point, a slower band centered at the Γ point dispersing towards ~0.1 eV below $E_F$ near the $\overline{X}$ point, and an electron pocket near the $\overline{X}$ point crossing $E_F$. The Fermi-level crossing (Fermi momentum $k_F$) of the electron pocket around the $\overline{X}$ point estimated by the fitting of the MDC at $E_F$ are ~0.70 and 1.26 ($\pi/a$). From comparison between the ARPES images taken at $h\nu = 82$ and 21.2 eV along with the Γ-$\overline{X}$-M line (Figs. 4(b) and 4(c)), the band dispersion changes with $h\nu$ due to the band dispersion along the $k_z$ direction. A preliminary comparison between the observation and the band calculation for $L1_0$-MnGa (see Appendix) suggests that the band dispersion observed by ARPES seemingly disagrees with the calculated band structure of $L1_0$-MnGa although a part of the observed band dispersion may agree with it. This implies that the $\sqrt{2} \times \sqrt{2}$ surface reconstruction affects the observed band dispersion and may modify the bulk band dispersion of $L1_0$-MnGa.

As mentioned above, the FSs around the $\overline{X}$ points (the corner of the rhombic FS) weekly depend on $k_z$. Figure 5(a) shows the ARPES image taken at $h\nu = 82$ eV along the $\overline{X}$-$\overline{M}$ line corresponding to the X-X line in the Brillouin zone of the primitive cell. We have observed a simple band structure, i.e., a hole-like band below $E_F$ and a shallow electron-like band crossing $E_F$ centered at the $\overline{X}$ point. These bands look degenerate at $E_B$ ~ 0.1 eV at the $\overline{X}$ point. Figures 5(b) and 5(c) show the second derivative plots of the ARPES images taken at $h\nu = 82$ and 21.2 eV, respectively. Since the electron band along the Γ-X̄-M line crosses $E_F$ with small $k_F$, as shown in Fig. 4(b), the electron band crossing $E_F$ along the X̄-M̄ line confirms the presence of the electron pocket around the X̄ point (see Fig. 2(b)).

We now focus on the band dispersion along the X̄-M̄ line around the X̄ point, namely, the shallow electron-like band crossing $E_F$. Figure 6(a) shows a magnified view of the X̄-M̄ band dispersion measured with $h\nu$ = 21.2 eV. Figure 6(b) shows the MDCs from $E_F$ to the binding energy $E_B$ = 0.1 eV. Each MDC is fitted by Lorentzian functions, as shown in Fig. 6(c), and their peak positions are plotted by open circles in Fig. 6(a). The Fermi velocity ($v_F$) is estimated as $5.8 \times 10^4$ (m/s) from the peak positions near $E_F$ by linear fitting (the green dashed line in Fig. 6(a)). This value is one order of magnitude smaller than that of alkali metals. Since the reciprocal of the MDC width at $E_F$ ($\Delta k_F$) is the mean free path $l$ of the conduction electron, i.e., $l = \frac{1}{\Delta k_F}$, we estimate the $l$ at $E_F$ to be 0.92 nm. This value is one order of magnitude smaller than that of elemental metals (~several tenth nm) [26]. Additionally, the curvature of the band dispersion corresponds with the effective mass $m^* = \frac{1}{\hbar^2}\frac{\partial E(k)}{\partial k \partial k} \sim 3.0 m_e$, where $m_e$ is the electron mass in a vacuum. The $m^*$ in MnGa is several times heavier than $m^*$ of ordinary metals (~$m_e$) and lighter than $m^*$ of transition metals, e.g., $m^* \sim 28\ m_e$ of Ni. The light $m^*$ of MnGa compared with typical ferromagnetic 3$d$ transition metals suggests that the renormalization of the band dispersion due to the electron correlation is not as strong as those metals. From these values, the mobility of the X̄-M̄ band $\mu_{\overline{XM}}$ is estimated to be ~9.23 (cm$^2$/Vs) using the equation $\mu_{\overline{XM}} = \frac{e\tau}{m^*} = \frac{el}{m^* v_F}$, where $e$ is the elementary charge and $\tau$ is the relaxation time. The value of $\mu_{\overline{XM}}$ is approximately one-third of the bulk one $\mu_{3D} = 28.4$ (cm$^2$/Vs) estimated from transport measurements. This result suggests that the X̄-M̄ band crossing $E_F$ is not very mobile.

The corners of the rhombic FSs around the $\bar{\text{X}}$ points forming the electron pocket may play a key role in recently observed giant interfacial spin-dependent properties in MnGa-based heterostructures. For example, although MnGa has a very strong PMA which does not favor the formation of topological spin textures such as magnetic skyrmions, topological Hall effect [27] (a signature of skyrmions) has been observed in MnGa/Pt, MnGa/Ta [28], and MnGa/BiSb [19] bilayers. The topological Hall effect was observed even under zero magnetic field in MnGa/BiSb, indicating stable ground-state skyrmions with a giant interfacial DMI at the MnGa/BiSb interface. Key factors for the interfacial DMI are the spin-orbit coupling of heavy elements like Bi and its interfacial hybridization with the magnetic MnGa layers. Since DMI is an indirect magnetic interaction between two spins mediated by intermediate electronic states [29,30], the existence of the electron pocket in the MnGa film may contribute to the hybridization between the interfacial state inducing such giant DMI in MnGa heterostructure. Considering the in-plane crystallographic relationship between MnGa(100)/BiSb(1$\bar{1}$0) interface [13] and the topological surface states on BiSb(1$\bar{1}$0) [31], the small electron pocket at the $\bar{\text{X}}$ point in the MnGa film (Fig. 2(b)) likely overlaps with a Dirac cone state in BiSb(1$\bar{1}$0) [31], resulting in the increase of the interfacial hybridization between the bands forming the electron pocket of MnGa and the topological surface state of BiSb.

It should be mentioned here that the $k_z$ broadening and the surface reconstruction seem to have affected the ultraviolet ARPES spectra fundamentally due to its surface sensitivity. To determine the entire bulk band structure, bulk sensitive soft x-ray ARPES measurements on $L1_0$ MnGa and detailed comparison with band-structure calculations are desirable.

**Conclusion**

In conclusion, we have performed angle-resolved photoemission spectroscopy (ARPES) on the $L1_0$-MnGa thin film and revealed its band structure. The obtained Fermi surface in the $k_x$-$k_y$ plane shows large rhombic features centered at the Γ point and cross each other near the $\bar{\text{X}}$ point of the $\sqrt{2} \times \sqrt{2}$-reconstructed surface Brillouin zone. The Fermi surface and constant-energy surface mappings in the out-of-plane $k_z$-$k_x$ space demonstrate that the observed band dispersion strongly depends on $k_z$. The disagreement between the ARPES image and the bands calculated by DFT for $L1_0$-MnGa implies that the surface reconstruction affects the band dispersion observed by ARPES. The value of $m^*$ in MnGa is lighter than the typical value in ferromagnetic $3d$ transition metals, probably because of the weak renormalization effect in MnGa. The mobility of the $\bar{\text{X}}$-$\bar{\text{M}}$ band around the $\bar{\text{X}}$ point forming the electron pocket is less than one-third of the bulk MnGa mobility. The existence of the small electron pocket in the MnGa film may explain the recently observed giant interfacial Dzyaloshinskii-Moriya interaction in MnGa-based heterostructures due to the increase of the interfacial hybridization between the layers.

**Acknowledgment**

The authors thank H. Katayama-Yoshida for enlightening discussion. This work was supported by JST-CREST (JPMJCR18T5 and JPMJCR18T1), Grants-in-Aid for Scientific Research (18K03484) from the Japan Society for the Promotion of Science (JSPS), Japan. This work was partially supported by the Spintronics Research Network of Japan (Spin-RNJ). This work at KEK-PF was performed under the approval of the

**Appendix: Comparison with DFT calculation**

The obtained band structure of the MnGa thin film is compared with the DFT calculations. Figure 7 shows comparison of the experimental band structure with the calculations. Here, the calculated band structures are shifted upward relative to the $E_F$ by 0.75 eV to reproduce the observations (however, because the electronegativity of Mn is close to that of Ga, the energy shift due to the non-stoichiometry ($Mn_{0.6}Ga_{0.4}$) may be smaller, ~0.3 eV at most). Figure 7(a) shows comparison of the FSM. The calculated bands form the circular FS, the rhombic FS, and the rounded square-like FS centered at the Γ point in the reconstructed surface Brillouin zone (the Γ-$\bar{X}$-$\bar{M}$ plane). Compared with the observation, the circular and outer square-like FSs of the calculated bands are missing in the observed FSM. Figure 7(b) shows comparison of the band dispersion along the Γ-$\bar{X}$-M line. The majority-spin bands (red color) seem to reproduce the observed band dispersion, while the minority-spin bands look absent in the observation. Thus, these comparisons suggest that some $\sqrt{2} \times \sqrt{2}$ structure may have to be assumed in order to archive agreement between the observed band structure and the calculation.

**Figure**

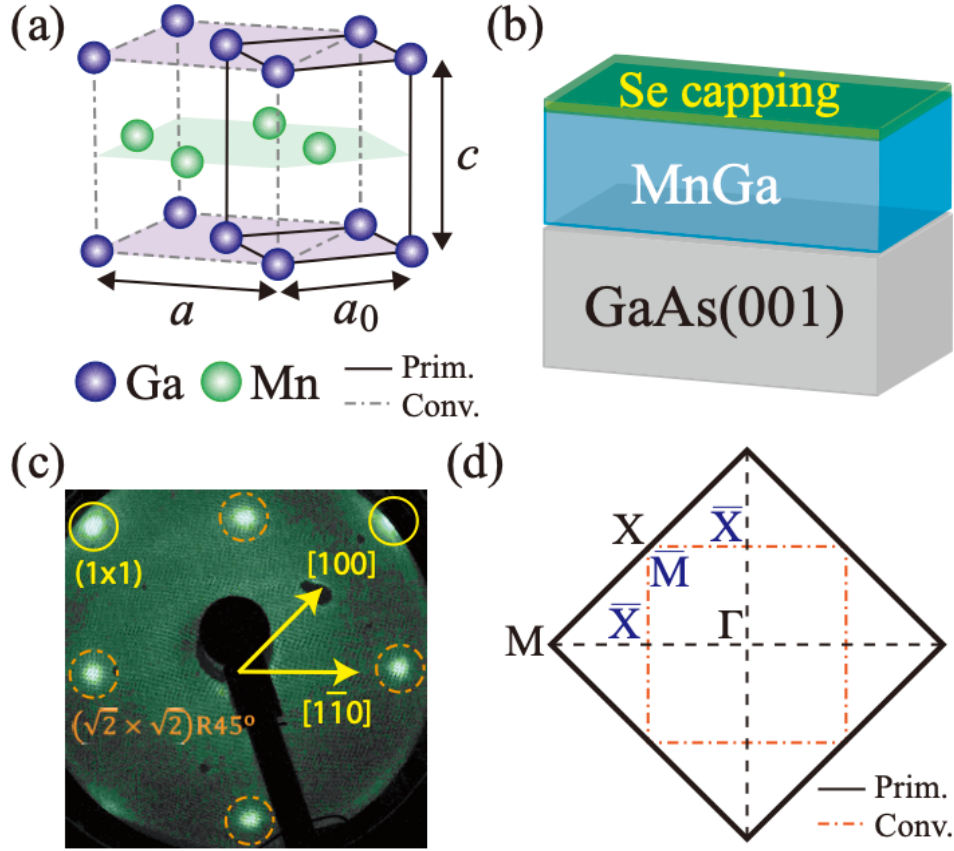

Fig. 1. Structure of $L1_0$-MnGa thin film. (a) $L1_0$ structure of MnGa. Solid and dashed lines denote the primitive cell and the conventional unit cell [9], respectively. $a$ $(=\sqrt{2}a_0)$ and $c$ are in-plane and out-of-plane lattice constants of the conventional unit cell, respectively. (b) Structure of the studied thin film. The amorphous Se capping layer protects the surface of the MnGa layer from oxidation. (c) LEED pattern of the MnGa layer after removing the amorphous Se capping layer by heating. Solid and dash-dotted circles denote diffraction spots from the $(1 \times 1)$ surface and the reconstructed $(\sqrt{2} \times \sqrt{2})R45°$ surface, respectively. (d) Brillouin zones of the MnGa thin film. Solid and dash-dotted lines denote the Brillouin zone boundaries in the Γ-X-M plane of the primitive and conventional cells, respectively. The letters with upper bars are symmetry points of the $(\sqrt{2} \times \sqrt{2})$ reconstructed Brillouin zone.

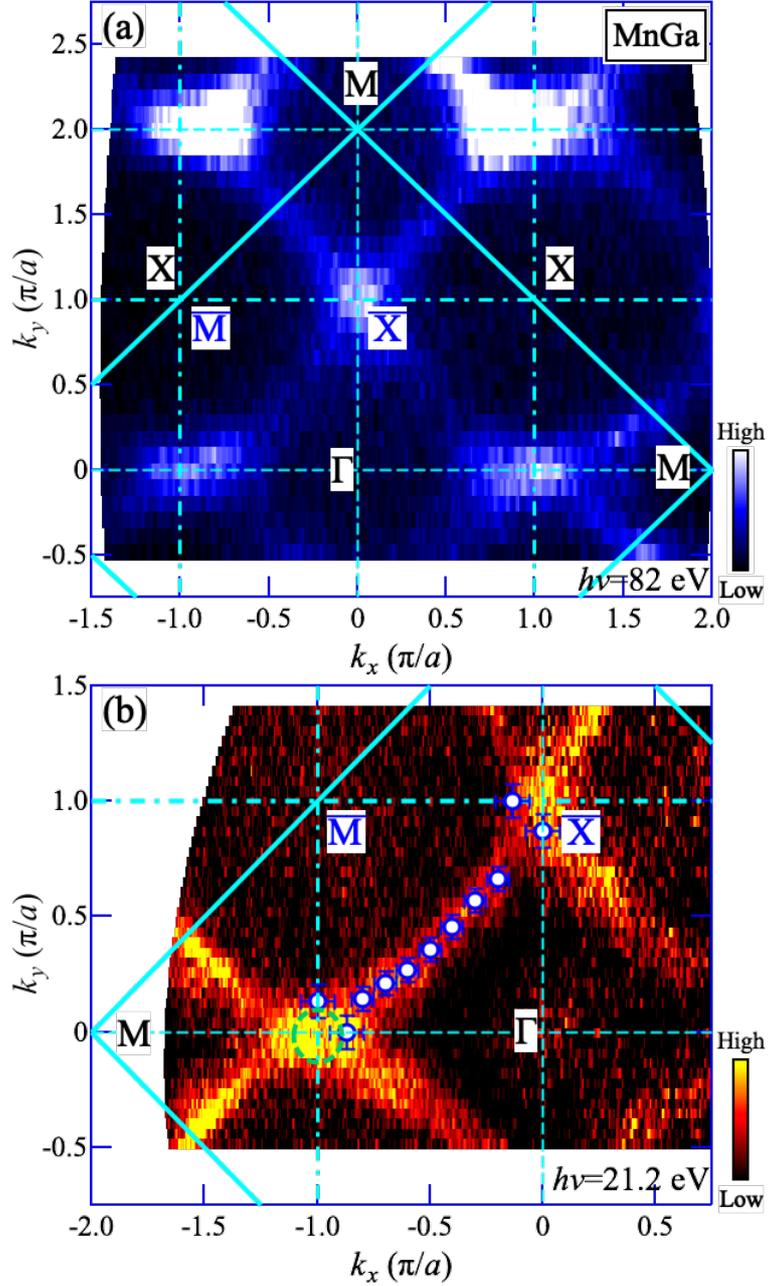

FIG. 2. Fermi surface mapping (FSM) of the MnGa thin film in the $k_x$ - $k_y$ space. (a) FSM taken at $h\nu = 82$ eV. Here, the surface normal momentum $k_z$ lies in the Γ-X-M plane. Solid and dot-dashed lines are the Brillouin zone boundaries of the primitive and conventional ($\sqrt{2} \times \sqrt{2}$-reconstructed) unit cells, respectively. Dashed lines represent symmetry crossing the zone centers. (b) FSM taken at $h\nu = 21.2$ eV. Here, the value of $k_z$ (~3.5 ($\pi/a$)) is not on the Γ-X-M plane. Open circles are the Fermi momenta estimated from the momentum distribution curves (MDCs) at $E_F$. Dashed circle around the $\bar{\text{X}}$ point denotes an electron pocket.

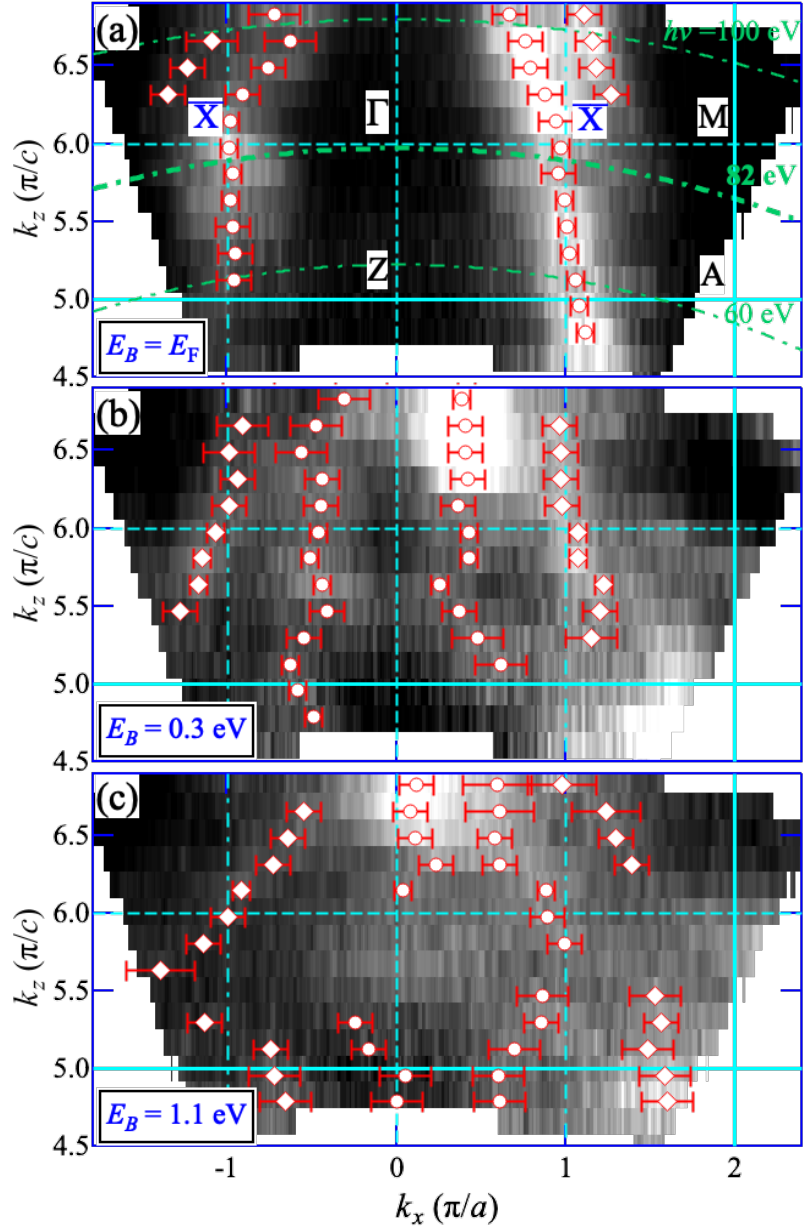

FIG. 3. Constant-energy surface mapping (CESM) of the MnGa thin film in $k_z - k_x$ space. (a) FSM in the $k_z$-$k_x$ plane. Green dash-dotted curves are **k**-space cuts for fixed $h\nu$'s. (b) $k_z$-$k_x$ CESM at binding energy $E_B = 0.3$ eV. (c) $k_z$-$k_x$ CESM at $E_B = 1.1$ eV. Solid and dot-dashed lines are the Brillouin zone boundaries of the primitive and conventional ($\sqrt{2} \times \sqrt{2}$-reconstructed) unit cells, respectively. Dashed lines are those crossing the zone centers. These mappings have been measured by varying $h\nu$. Open circles and rhombi are the peak positions of the MDCs estimated by Lorentzian fitting.

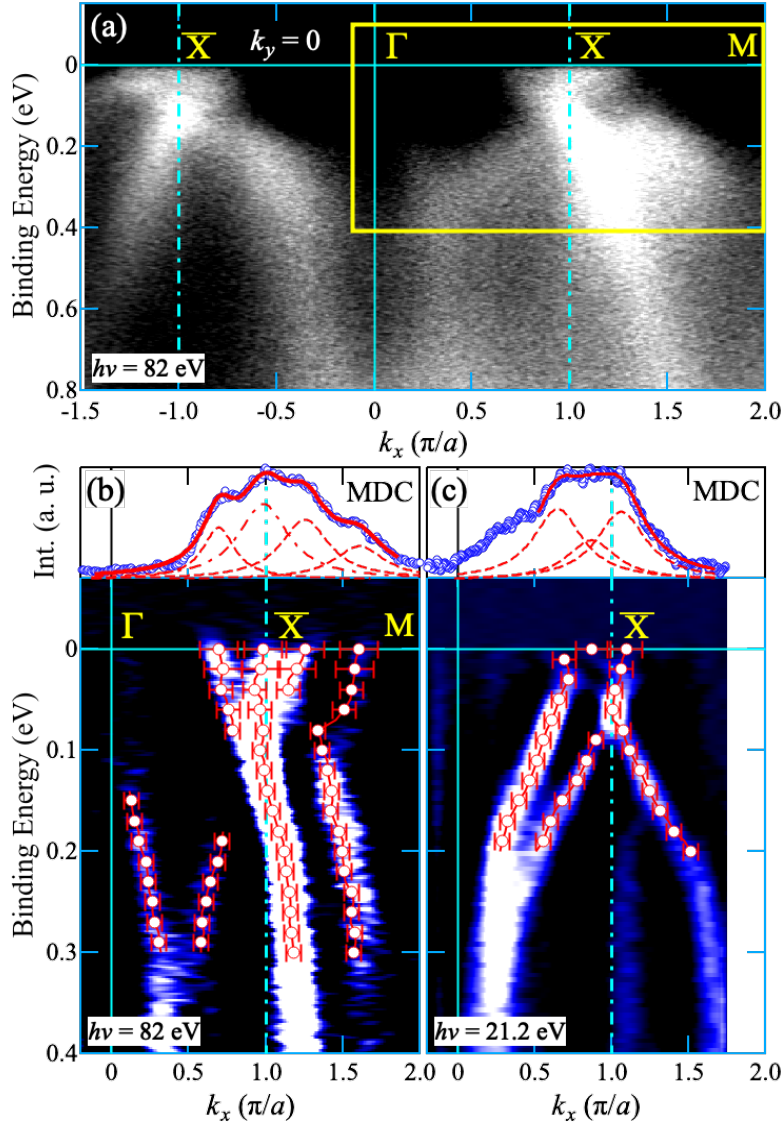

FIG. 4. Band dispersion along the Γ-$\overline{X}$ symmetry line of the MnGa thin film. (a) ARPES images taken at $h\nu$ = 82 eV along the Γ-$\overline{X}$ line. Squares represent the areas corresponding to panels (b) and (c). (b),(c) Second derivatives of the ARPES intensities taken at $h\nu$ = 82 eV and 21.2 eV, respectively. Open circles are the peak positions of momentum distribution curves (MDCs) estimated by Lorentzian fitting. Dot-dashed lines are the reconstructed Brillouin zone boundaries for the conventional ($\sqrt{2} \times \sqrt{2}$-reconstructed) unit cell. The top figures show MDCs at $E_F$.

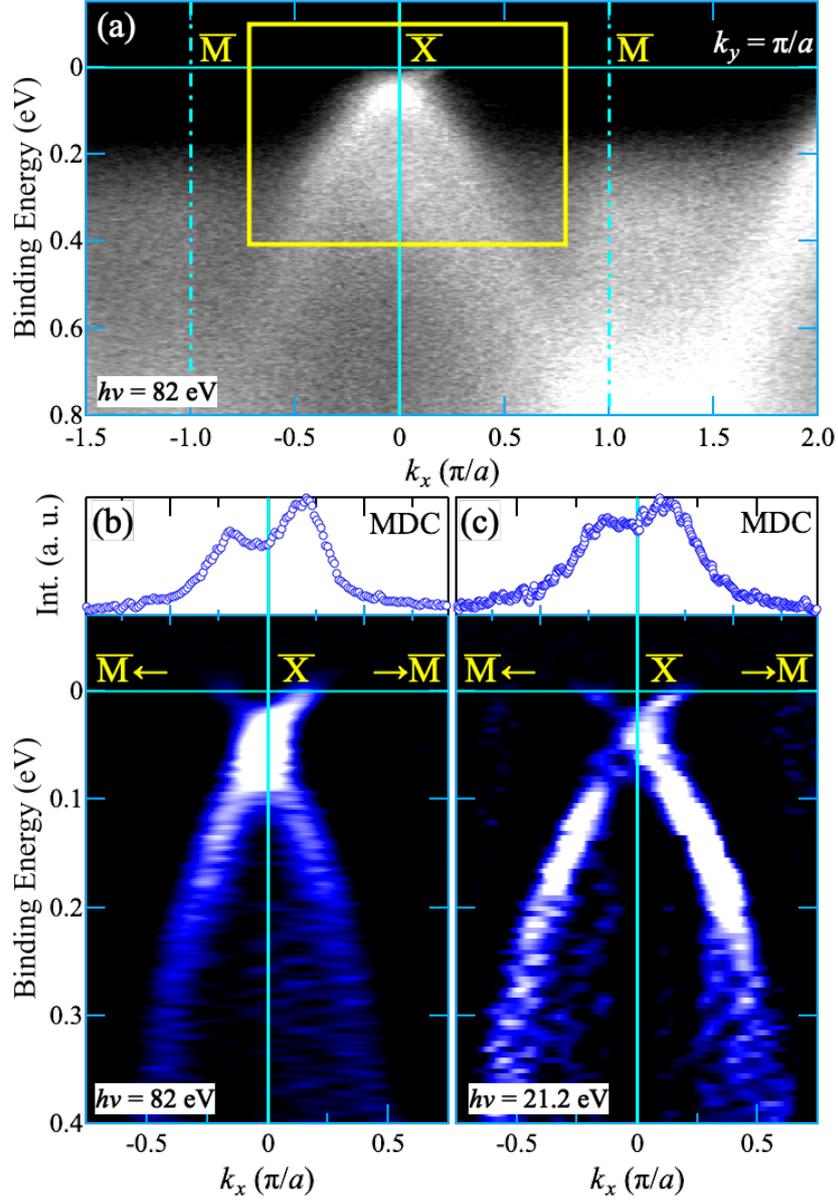

FIG. 5. Band dispersion along the $\overline{\text{X}}$-$\overline{\text{M}}$ symmetry lines of the MnGa thin film. (a) ARPES images taken at $h\nu = 82$ eV along the $\overline{\text{X}}$-$\overline{\text{M}}$ line. (b),(c) Second derivatives of the ARPES intensities taken at $h\nu = 82$ eV and 21.2 eV, respectively. The top figures show MDCs at $E_\text{F}$.

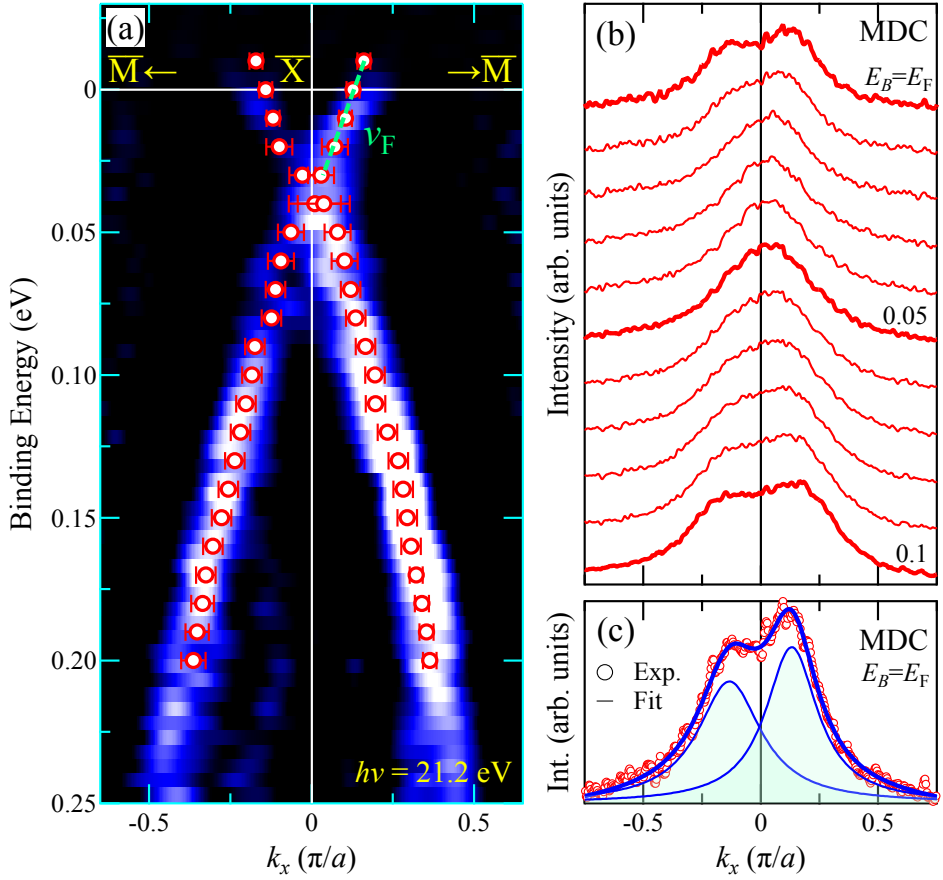

FIG. 6. Band dispersion near the $\overline{X}$ point in MnGa. (a) ARPES images along the $\overline{X}$-$\overline{M}$ symmetry lines. The green dashed line denotes the slope of the band dispersion near $E_F$ corresponding to the Fermi velocity ($v_F$). (b) MDCs from $E_F$ to $E_B = 0.1$ eV. (c) Lorentzian fitting for the MDC at $E_B = E_F$. Open circles in panel (a) are the peak positions of MDCs estimated by the fitting.

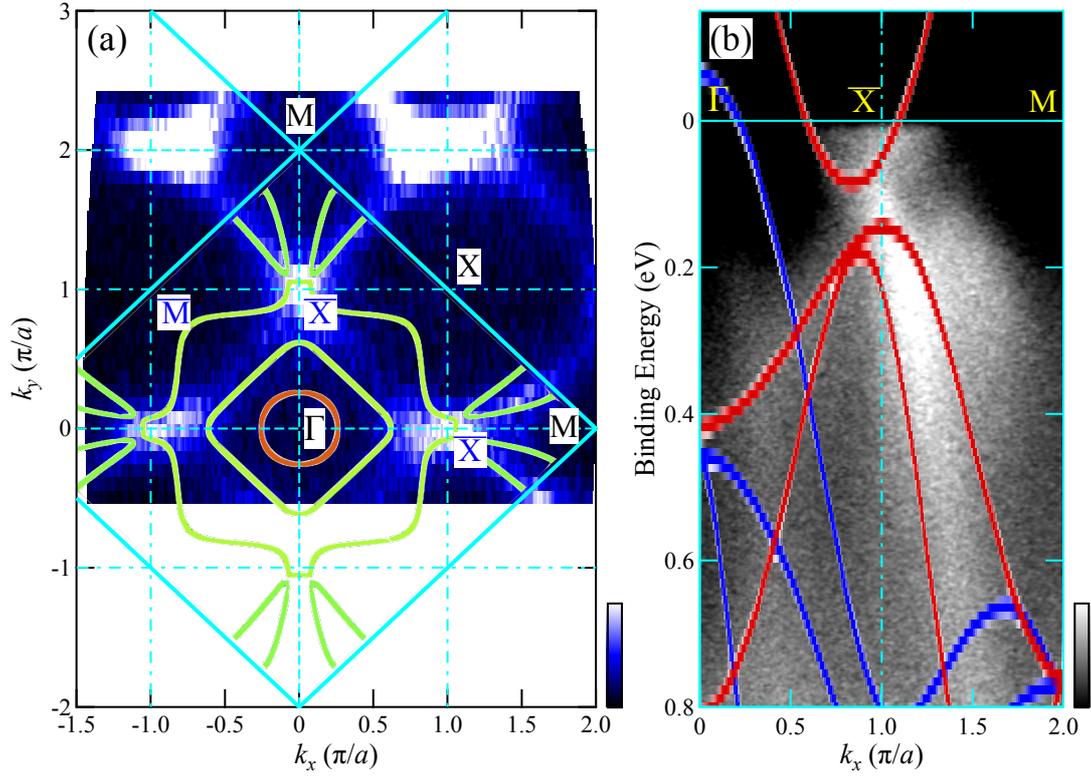

FIG. 7. Comparison of the experimental band structure with the DFT calculation. (a) Fermi surface mapping. (b) Band dispersion along the $\Gamma$-$\overline{\mathrm{X}}$ symmetry line. The red and blue solid lines are the calculated bands for the majority-spin and minority-spin states, respectively.

**Reference**

[1] K. M. Krishnan, Appl. Phys. Lett. **61**, 2365 (1992).

[2] M. Tanaka, J. P. Harbison, J. DeBoeck, T. Sands, B. Philips, T. L. Cheeks, and V. G. Keramidas, App. Phys. Lett. **62**, 1565 (1993).

[3] L. J. Zhu, S. H. Nie, and J. H. Zhao, Chin. Phys. B **22**, 118505 (2013).

[4] S. Zhao, and T. Suzuki, AIP Adv. **6**, 056025 (2016).

[5] W. Y. Zhang, P. Kharel, S. Valloppilly, R. Skomski, and D. J. Sellmyer, J. Appl. Phys. **117**, 17E306 (2015).

[6] K. Wang, E. Lu, J. W. Knepper, F. Yang, and A. R. Smith, Appl. Phys. Lett. **98**, 162507 (2011).

[7] S. Mizukami, T. Kubota, F. Wu, X. Zhang, T. Miyazaki, H. Naganuma, M. Oogane, A. Sakuma, and Y. Ando, Phys. Rev. B **85**, 014416 (2012).

[8] F. Wu, E. P. Sajitha, S. Mizukami, D. Watanabe, T. Miyazaki, H. Naganuma, M. Oogane, and Y. Ando, Appl. Phys. Lett. **96**, 042505 (2010).

[9] M. Tanaka, Mater. Sci. Eng. B **31**, 117 (1995).

[10] A. Köhler, I. Knez, D. Ebke, C. Felser, and S. S. P. Parkin, Appl. Phys. Lett. **103**, 162406 (2013).

[11] J. N. Feng, W. Liu, W. J. Gong, X. G. Zhao, D. Kim, C. J. Choi and Z. D. Zhang, J. Mater. Sci. Tech. **33**, 291 (2017).

[12] L. Zhu, S. Nie, K. Meng, D. Pan, J. Zhao, and H. Zheng, Adv. Mater. **24**, 4547 (2012).

[13] N. H. D. Khang, Y. Ueda, K. Yao, and P. N. Hai, J. Appl. Phys. **122**, 143903 (2017).

[14] S. Mizukami, F. Wu, A. Sakuma, J. Walowski, D. Watanabe, T. Kubota, X. Zhang, H. Naganuma, M. Oogane, Y. Ando, and T. Miyazaki, Phys. Rev. Lett. **106**, 117201 (2011).

[15] S. Mao *et al.*, Sci. Rep. **7**, 43064 (2017).

[16] K. Chonan, N. H. D. Khang, M. Tanaka, P. N. Hai, Jpn. J. Appl. Phys. **59**, SGGI08 (2020).

[17] R. Ranjbar, K. Z. Suzuki, Y. Sasaki, L. Bainsla, and S. Mizukami, Jpn. J. Appl. Phys. **55**, 120302 (2016).

[18] N. H. D. Khang, Y. Ueda, P. N. Hai, Nat. Mater. **17**, 808 (2018).

[19] N. H. D. Khang, T. Fan, and P. N. Hai, AIP Advances **9**, 125309 (2019).

[20] S. Mizukami, T. Kubota, F. Wu, X. Zhang, T. Miyazaki, H. Naganuma, M. Oogane, A. Sakuma, and Y. Ando, Phys. Rev. B **85**, 014416 (2012).

[21] Z. Yang, J. Li, D.-S. Wang, K. Zhang, and X. Xie, J. Magn. Magn. Mater. **182**, 369 (1998).

[22] G. Kresse and J. Furthmüller, Phys. Rev. B **54**, 11169 (1996)

[23] J. P. Perdew, K. Burke, and M. Ernzerhof, Phys. Rev. Lett. **77**, 3865 (1996).

[24] T. Grandke, L. Ley, and M. Cardona, Phys. Rev. B **18**, 3847 (1978).

[25] T. Mitsuhashi, M. Minohara, R. Yukawa, M. Kitamura, K. Horiba, M. Kobayashi, and H. Kumigashira, Phys. Rev. B **94**, 125148 (2016).

[26] D. Gall, J. Appl. Phys. **119**, 085101 (2016).

[27] P. Bruno, V. K. Dugaev, and M. Taillefumier, Phys. Rev. Lett. **93**, 096806 (2004).

[28] K. K. Meng, X. P. Zhao, P. F. Liu, Q. Liu, Y. Wu, Z. P. Li, J. K. Chen, J. Miao, X. G. Xu, J. H. Zhao, and Y. Jiang, Phys. Rev. B **97**, 060407(R) (2018).

[29] I. Dzyaloshinsky, J. Phys. Chem. Solids **4**, 241 (1958).

[30] T. Moriya, Phys. Rev. **120**, 91 (1960).

[31] X.-G. Zhu, M. Stensgaard, L. Barreto, W. Simoes e Silva, S. Ulstrup, M. Michiardi, M. Bianchi, M. Dendzik, and P. Hofmann, New J. Phys. **15**, 103011 (2013).